\documentclass[10pt,a4paper]{article}

\usepackage{amsmath,amsfonts,amsthm}
\usepackage{graphicx}
\usepackage{multicol}
\usepackage{graphicx,times}             
\usepackage[misc]{ifsym}
\usepackage{cite}
\usepackage{stmaryrd}
\usepackage{mathtools}
\usepackage{lineno,hyperref}
\usepackage{latexsym}
\usepackage{caption}

\usepackage[top=1cm, bottom=2.29cm, left=1.6cm, right=1.47cm]{geometry}
\usepackage{sectsty}
\sectionfont{\normalsize}
\makeatletter
   \def\@seccntformat#1{\csname the#1\endcsname.\quad}
\makeatother

\usepackage{titlesec}
\titlespacing{\section}{0pt}{6pt}{3pt}

\begin{document}

\begin{center}
\begin{huge}\textbf{
	Scalar field and time varying Cosmological constant in $f(R,T)$ gravity for Bianchi type-I Universe
	}\end{huge}\\[10pt]
\textbf{G. P. Singh$^{1}$\footnote{gpsingh@mth.vnit.ac.in (GPS)}, Binaya K. Bishi$^{1}$\footnote{binaybc@gmail.com (BKB)}, P. K. Sahoo$^2$\footnote{sahoomaku@rediffmail.com (PKS)}}\\
\textbf{Department of Mathematics, Visvesvaraya National Institute of Technology,
Nagpur-440010, India$^{1}$}\\
\textbf{Department of Mathematics, Birla Institute of Technology and Science-Pilani, Hyderabad Campus, Hyderabad-500078, India$^{2}$}\\
\end{center}

\begin{abstract}
In this article, we have analysed the behaviour of scalar field  and cosmological constant in $f(R,T)$ theory of gravity. Here, we have considered the simplest form  of $f(R,T)$ i.e. $f(R,T)=R+2f(T)$, where $R$ is the Ricci scalar and $T$ is the trace of the energy momentum tensor and explored the spatially homogeneous and anisotropic Locally Rotationally Symmetric (LRS) Bianchi type-I cosmological model. It is assumed that the Universe is filled with two non-interacting matter sources namely scalar field (normal or phantom) with scalar potential and matter contribution due to $f(R,T)$ action. We have discussed two cosmological models according to power law and exponential law of the volume expansion along with constant and exponential scalar potential as sub models. Power law models are compatible with normal (quintessence) and phantom scalar field whereas exponential volume expansion models are compatible with only normal (quintessence) scalar field. The values of cosmological constant in our models are in agreement with the observational results. Finally, we have discussed some physical and kinematical properties of both the models.
\end{abstract}

\section{Introduction}
In this modern era, it is strongly believed that we are living in an accelerating Universe. It has been established through different observations like Ia supernova \cite{Riess98,Perlmutter99,Suzuki12}, Cosmic Microwave Background (CMB) anisotropy from Planck \cite{Ade14}, Wilkinson Microwave Anisotropy Probe (WMAP) \cite{Hinshaw13}, Large Scale Structures (LSS) \cite{Spergel03}, Baryon Acoustic Oscillations (BAO) \cite{Anderson13}. These observations have also pointed out that the Universe is dominated by some kind of exotic matter, known as dark energy \cite{Anderson13,Sahni04,Frieman08}. It is supposed that the dark energy was responsible to produce sufficient acceleration in late time evolution of the Universe. Thus, it is much more essential to study the fundamental nature of the dark energy, and several attempts have been made to understand it. The cosmological constant is assumed to be the simplest candidate of dark energy. It is the classical correction made to the Einstein's field equations by adding cosmological constant to the field equations. The introduction of cosmological constant to Einstein's field equations is the most efficient way of generating accelerated expansion. Later it faces serious problems like fine-tuning and cosmic coincidence problems in cosmology \cite{Peebles03,Sahni00}. There are also other claimants of dark energy like quintessence \cite{Martin08}, phantom \cite{Nojiri06}, k-essence \cite{Chiba00}, tachyons \cite{Padmanabhan02}, Chaplygin gas \cite{Bento02}. However, there is no direct detection of such exotic fluids. Quintessence is considered as the fifth element of the Universe along with four other elements namely baryons, radiations, dark matter and neutrinos. The name "Quintessence"  has been introduced by Caldwell et al. \cite{Caldwell98}. The quintessence field arises from Lagrangian, which is motivated by relativistic continuum mechanics. It is assumed that the equation of state for quintessence is $p_e=\omega_e\rho_e$, where $\rho_e$ and $p_e$ are the energy density and pressure respectively. Research indicates that observational constraint in the equation of state of dark energy suggests $1.38 < \omega_e < 0.82$ \cite{Melchiorri03}. The phantom is  one of the candidature of dark energy, which is a scalar field with negative kinetic energy. In this case, the equation state parameter satisfies the condition  $\omega_e<-1 $ \cite{Johri04}. It is known that for $\omega_e<-1 $, the dark energy (phantom) will violate all energy conditions \cite{Wald84}. Still, such type of models \cite{Caldwell02} are consistent with observational data $(1.46 < \omega_e < 0.78)$ \cite{Knop03,Riess04}. It is important to note that in case of phantom model, the Universe will end with a big rip. In other words, for a phantom dominated Universe, its total life time is finite \cite{RongGen05}.
\par
K-essence is a simple way for constructing an accelerated expansion model of the Universe. It is exclusively based on the idea that the unknown dark energy component is due to a minimally coupled scalar field with non-canonical kinetic energy which results in the negative pressure and provides K-essence cosmological model \cite{Putter07}. Tachyon is also  another candidature for dark energy. Such type of models are based on scalar field. The tachyon dark energy has an equation of state parameter between 1 and 0. The tachyon is an unstable field, which has become important in the string theory through its role in the Dirac-Born-Infield action, which is used to describe the D-brane action \cite{Sadeghi15}. Chaplygin gas is also one of the prominent candidature of dark energy. It is a perfect fluid satisfying the equation of state $p=\frac{-A}{\rho}$, where $A$ is a positive constant. This equation of state has been introduced by Chaplygin \cite{Chaplygin04} as a suitable mathematical approximation for calculating the lifting force on a wing of an airplane in aerodynamics \cite{Gorini04}.  Modification is made to the equation of state of Chapligin gas and it has been named as Generalized  Chapligin gas, Modified Chapligin gas etc. Due to the absence of strong evidence of existence about the dark energy, many researchers are interested in exploring the dark energy. Bamba et al.\cite{Bamba12} have studied the number of dark energy models like  $\Lambda$CDM model, Little Rip and Pseudo-Rip scenarios, the phantom and quintessence cosmologies with the four types (I, II, III and IV) of the finite-time future singularities and non-singular universes filled with dark energy.
 \par
One possible way of exploring dark energy is to modify the geometric part of the Einstein-Hilbert
action \cite{Magnano87}. Based on its modifications, several alternative theories of gravity came into existence. Some of the modified theories of gravity are $f(R)$, $f(T)$, $f(G)$ and $f(R,T)$ gravity. These models were proposed to explore the dark energy and other cosmological problems.  The $f(R)$ modified theory produces both cosmic inflation and mimic behavior of dark energy including  present cosmic acceleration \cite{DeFelice05,Sotiriou10,Nojiri11}. Amendola et al. \cite{Amendola07} have discussed the cosmologically viable conditions in $f(R)$ theory, which describes the dark energy models. Jamil et al. \cite{Jamil12} have discussed Reconstruction of some cosmological models in $f(R,T)$ cosmology.
\par
Scalar field cosmology plays a crucial role in the study of the early Universe, particularly in the investigation of inflation \cite{Guth81,Olive90}. It acts as the most natural basis for the inflationary cosmology. Thus, scalar field cosmology gained popularity among researchers. Research suggests that most of the studies have been carried out with a minimally coupled scalar field representing quintessence.  Ellis and Madsen \cite{Ellis91} have studied the FRW model with a minimally coupled scalar field along with a potential and a perfect fluid in the form of radiation. Barrow and Saich \cite{Barrow93} have studied the class of exact isotropic cosmological models containing perfect fluids and scalar fields with nonzero potentials and discussed the conditions under which inflation can occur and when the scalar field source dominates the late stages of the expansion of the Universe. Non-minimally coupled scalar field cosmological models have been discussed by Barrow and Mimoso \cite{Barrow94} and  Mimoso  and Wands \cite{Mimoso95}. Cosmological models involving perfect fluid and scalar field with an exponential potential have been thoroughly analysed \cite{Billyard99}. Singh and Singh \cite{Singh11} have discussed the FRW models with perfect fluid and scalar field in higher derivative theory. Reconstruction of scalar field dark energy models in kaluza-klein Universe are studied by Sharif and Jawad \cite{Sharif13}. Sharif and Zubair\cite{Sharif12} have investigated the Anisotropic Universe Models with Perfect Fluid and Scalar Field in  $f(R,T)$ Gravity. Recently Singh and Singh \cite{Singh14} have explored the behaviors of scalar field in modified $f(R,T)$ gravity for  flat Friedmann-Robertson-Walker cosmological model.
\par
A lot of investigations have been carried out by several authors \cite{SahooSivakumar15,singhbishi15a,singhbishi15b,singhbishi15c,Sahoomishra14,mishraSahoo14,Sahooetal14} in $f(R,T)$ modified gravity.  Sahoo and  Sivakumar \cite{SahooSivakumar15} have discussed the "LRS Bianchi type-I cosmological model in $f(R,T)$ theory of gravity with $\Lambda(T)$. In this article they have presented the second class of $f(R,T)$ modify theory of gravity for the LRS Bianchi type-I space time and obtained the exact solutions of the modified Einstein’s field equations with the help of linearly varying deceleration parameter. Also perfect fluid form of energy momentum tensor is considered here.
Singh and Bishi \cite{singhbishi15a,singhbishi15b}  have invested the first and second class of of $f((R,T)$ gravity for the anisotropic Bianchi type-I space time.  Expansion law (polynomial and exponential) is used to solve the field equations  in Singh and Bishi \cite{singhbishi15b}, where as generalized form of scale factor is used in Singh and Bishi \cite{singhbishi15a}. In fact Singh and Bishi \cite{singhbishi15b}, is generalized form of Singh and Bishi \cite{singhbishi15a}.  Quadratic form of energy momentum tensor is considered in both the articles.
Singh and Bishi \cite{singhbishi15c} is focused on the study of FRW space time for a variable gravitational and cosmological constant with the modified Chaplygin gas equation of state in $f(R,T)$ gravity.  In this article Hybrid exponential law is used to obtain exact solutions of the field equations.
Sahoo and  Mishra \cite{Sahoomishra14} have investigated the  Kaluza–Klein dark energy model in the form of wet dark fluid in $f(R,T)$ gravity.  In this article they have explored the first type of $f(R,T)$ gravity in presence of wet dark fluid for five dimensional Kaluza-Klein space time and also used expansion law (polynomial and exponential) to solve the modified Einstein’s field equations.
Bianchi type ${VI}_h$ cosmological model for the first type of $f(R,T)$ gravity is analysed by the same authors \cite{mishraSahoo14}.  They have used the Berman’s law i.e. expansion scalar is proportional to the shear scalar to obtain the exact solutions of Einstein’s modified field equations for $h=1$ and $h= -1$ case. Sahoo et al. \cite{Sahooetal14} have discussed the first type of $f(R,T)$ gravity for uniform, anisotropic and axially symmetric line element in presence of prefect fluid.  The variation law of mean Hubble parameter is used to obtain the exact solutions of modified Einstein’s field equations and some physical and kinematical properties of the model are also discussed.
It is hoped that the articles
\cite{SahooSivakumar15,singhbishi15a,singhbishi15b,singhbishi15c,Sahoomishra14,mishraSahoo14,Sahooetal14} and other discussed articles in the above contributes to the investigation of $f(R,T)$ gravity, which is useful for the astrophysicists and cosmologists of the different corner of the world.
\par
Motivated by the above mentioned research, in this present article we have investigated the first form of $f(R,T)$ gravity for LRS Bianchi type-I space time in presence of scalar field and cosmological constant.  Constant scalar potential and exponential scalar potential with expansion law (Polynomial and exponential) is used to obtain the exact solution of the modified Einstein’s field equations. The main difference between the past articles \cite{SahooSivakumar15,singhbishi15a,singhbishi15b,singhbishi15c,Sahoomishra14,mishraSahoo14,Sahooetal14} and this article is that they differ from each other in terms of matter source and the solution procedure to obtain the exact solution of the field equations.

\section{Gravitational field equations of $f(R,T)$ gravity with scalar field}
The field equations of $f(R,T)$ gravity with minimally couples scalar field $\phi$ in a gravitational action self interacting through potential $V(\phi)$ is given as \cite{Singh14}
\begin{equation}
\label{e1}
S=\frac{1}{2}\int {[f(R,T)+2\mathcal{L}_{\phi}]\sqrt{-g}d^{4}x}
\end{equation}
where $f(R,T)$ is an arbitrary function of Ricci scalar $(R)$, and $T$
is the trace of energy-momentum tensor $(T_{ij})$ of the matter, $\mathcal{L}_{\phi}$
is the matter Lagrangian of scalar field. The energy momentum tensor
$T_{ij}$ of matter is defined as
\begin{equation}
\label{e2}
T_{ij}= - \frac{2}{\sqrt{-g}} \frac{\delta(\sqrt{-g}\mathcal{L}_{\phi})}{\delta
g^{ij}}
\end{equation}
and its trace by $T=g^{ij}T_{ij}$. The matter Lagrangian $\mathcal{L}_{\phi}$ depends only on the metric tensor $g_{ij}$
rather than its derivatives. Hence, equation \eqref{e2} gets reduced to
\begin{equation}
\label{e3}
T_{ij}= g_{ij}\mathcal{L}_{\phi} - 2\frac{\partial \mathcal{L}_{\phi}}{\partial g^{ij}}
\end{equation}
The field equations of $f(R,T)$ gravity by varying the
action $S$ with respect to metric tensor $g_{ij}$ can be written as
\begin{eqnarray}
\label{e4}
f_R(R,T)R_{ij}-\frac{1}{2} f(R,T) g_{ij} +
(g_{ij}\Box-\nabla_i\nabla_j)f_R(R,T)=T_{ij}-f_T(R,T) T_{ij}
-f_T(R,T) \Theta_{ij}
\end{eqnarray}
where, $f_R(R,T)$ and $f_T(R,T)$ are the derivatives partially with respect to $R$ and $T$ respectively.
Here, $\nabla_i$ denotes the usual covariant derivative and $\Theta_{ij}$ is defined as
\begin{equation}
\label{e6}
\Theta_{ij}= \frac{g^{ij}\delta T_{ij}}{\delta g^{ij}}=-2T_{ij}+g_{ij}\mathcal{L}_{\phi} - 2g^{lm}
\frac{\partial^2 \mathcal{L}_{\phi}}{\partial g^{ij} \partial g^{lm}}
\end{equation}
$\Box\equiv\nabla^i\nabla_i$ is the D'Alembert operator which is defined as
\begin{equation}
\label{e5}
\Box=\frac{1}{\sqrt{-g}}\partial_{i}(\sqrt{-g}g^{ij})\partial_{j}
\end{equation}
Using contraction of indices one can rewrite  \eqref{e4} as
\begin{eqnarray}
\label{e7}
Rf_R(R,T)+ 3\Box f_R(R,T)- 2f(R,T)=T-Tf_T(R,T) -f_T(R,T) \Theta
\end{eqnarray}
where $\Theta=\Theta^i_i$.
Since the field equations of $f(R,T)$ gravity depend on the physical nature of $\Theta_{ij}$, so depending on the nature of the matter source, one can obtain several theoretical models corresponding to different matter
contributions for $f(R,T)$ gravity. Harko et al.\cite{Harko11} presented three classes of these models as follows
\begin{equation}
\label{e8}
f(R,T )=
\left\{
\begin{array}{lcl}
    R+2f(T) \\
    f_1(R)+f_2(T) \\
    f_1(R)+f_2(R)f_3(T)
\end{array}
\right.
\end{equation}
In this paper, let us assume the simplest form of $f$ as $f(R,T)=R+2f(T)$, where $R$ is a function of cosmic time and $2f(T)$ is the gravitational interaction between matter and curvature. Using this into the field equations (4), we can get the field equations of $f(R,T)$ gravity with cosmological term $\Lambda$ as
\begin{equation}
\label{e9}
R_{ij}-\frac{1}{2}Rg_{ij}=T_{ij}-2(T_{ij}+\Theta_{ij})h^{\prime}(T)+h(T)g_{ij}+\Lambda g_{ij}
\end{equation}
Assuming the Universe is filled with scalar field coupled to gravity, the energy-momentum tensor of a scalar field $\phi$ with self-interacting scalar field potential $V(\phi)$ is given by
\begin{equation}
\label{e10}
T_{ij}=\epsilon\partial_i\phi\partial_j\phi-\biggl[\frac{\epsilon}{2}\partial_l\phi\partial^l\phi-V(\phi)\biggr]g_{ij}
\end{equation}
where $\epsilon=\pm1$ corresponding to normal (quintessence) and phantom scalar fields, respectively. Then the matter Lagrangian of the scalar field can be taken as
\begin{equation}
\label{e11}
\mathcal{L}_{\phi}=-\frac{1}{2}\epsilon \phi^{\prime2}+V(\phi)
\end{equation}
where, an overhead prime denotes ordinary derivative with respect to cosmic time $"t"$.
Using this in \eqref{e6}, we get
\begin{equation}
\label{e12}
\Theta_{ij}= -2T_{ij}-\biggl(\frac{1}{2}\epsilon \phi^{\prime2}-V(\phi)\biggr)g_{ij}
\end{equation}
The trace of the energy momentum tensor for our model is given by
\begin{equation}
\label{e13}
T=-\epsilon\phi^{\prime2}+4V(\phi)
\end{equation}
Throughout the paper the units are taken such that $G=c=1$, where $G$ is the Newtonian gravitational constant and $c$ is the speed of light in free space.

\section{Metric and Field Equations}

We consider the spatially homogeneous and anisotropic LRS Bianchi type-I metric of the form
\begin{equation}
\label{e14}
ds^{2}=dt^{2}-A^{2}(dx^{2}+dy^{2})-B^{2}dz^{2}
\end{equation}
where $A$ and $B$ are functions of cosmic time $t$ only. The eccentricity of the above Universe is given by $e=\sqrt{1-\frac{B^2}{A^2}}$. The directional Hubble parameters are $H_1=\frac{A^{\prime}}{A}$ and $H_3=\frac{B^{\prime}}{B}$, hence the mean Hubble parameter $H=\frac{2H_1+H_3}{3}$ and $\theta=3H$. The average scale factor, spatial volume, scalar expansion for this metric \eqref{e14} are obtained respectively as
\begin{equation}
\label{e15}
a=(A^2B)^{\frac{1}{3}},\ \ \mathbf{V}=a^3=A^2B, \ \ \theta=u^i_{;i}=2H_1+H_3
\end{equation}
The shear scalar is defined as
\begin{equation}
\label{e16}
\sigma^2=\frac{1}{2} \biggl[\sum H^2_i-\frac{1}{3}\theta^2\biggr]=\frac{1}{3}(H_1-H_3)^2
\end{equation}
For the choice of $h(T)=\lambda T$, where $\lambda$ is an arbitrary constant,
the field equations \eqref{e9} for the line element \eqref{e14} with the help of \eqref{e15} takes the form
\begin{equation}
\label{e17}
H_1^2+2H_1H_3=-\biggl(\frac{2\lambda+1}{2}\biggr)\epsilon(\phi^{\prime})^2-(4\lambda+1)V(\phi)-\Lambda(t)
\end{equation}
\begin{equation}
\label{e18}
H_1^{\prime}+H_3^{\prime}+H_1^2+H_3^2+H_1H_3=\frac{2\lambda+1}{2}\epsilon(\phi^{\prime})^2-(4\lambda+1)V(\phi)-\Lambda(t)
\end{equation}
\begin{equation}
\label{e19}
2H_1^{\prime}+3H_1^2=\frac{2\lambda+1}{2}\epsilon(\phi^{\prime})^2-(4\lambda+1)V(\phi)-\Lambda(t)
\end{equation}
From \eqref{e17} and \eqref{e18},
\begin{equation}
\label{e20}
\Lambda(t)=\frac{-1}{2}\left[(8\lambda+2)V(\phi)+H_1^{\prime}+H_3^{\prime}+2H_1^2+H_3^2+3H_1H_3\right]
\end{equation}
Using the value of $\Lambda$ in \eqref{e17}, we get
\begin{equation}
\label{e21}
(\phi^{\prime})^2=\frac{1}{(2\lambda+1)\epsilon}\left[H_1^{\prime}+H_3^{\prime}+H_3^2-H_1H_3 \right]
\end{equation}
Using \eqref{e15} in \eqref{e17}-\eqref{e19}, we obtain
\begin{equation}
\label{e22}
A=\mathbf{V}^{\frac{1}{3}}l_2^{\frac{1}{3}}exp\left(\frac{l_1}{3}\int \frac{1}{\mathbf{V}}dt\right)
\end{equation}
and
\begin{equation}
\label{e23}
B=\mathbf{V}^{\frac{1}{3}}l_2^{\frac{-2}{3}}exp\left(\frac{-2l_1}{3}\int \frac{1}{\mathbf{V}}dt\right)
\end{equation}
\section{Power Law Volume Expansion Model }

Since equations \eqref{e17}-\eqref{e19} are three equations in five unknowns $A, B, \phi, V(\phi), \Lambda$, it require some additional conditions to get a cosmological model with proper solutions. Here, we have considered two cases for scalar potential  $V(\phi)$  and the power law expansion of the volume
\begin{equation}
\label{e24}
\mathbf{V}=t^{3n}
\end{equation}
where $n$ is a constant. Using the above in \eqref{e22} and \eqref{e23}, we get the scale factors as

\begin{equation}
\label{e25}
A=l_2^{\frac{1}{3}}t^ne^{\frac{-l_1t^{-3n+1}}{3(3n-1)}}
\end{equation}
and
\begin{equation}
\label{e26}
B=l_2^{\frac{-2}{3}}t^ne^{\frac{2l_1t^{-3n+1}}{3(3n-1)}}
\end{equation}
The directional Hubble parameters are
\begin{equation}
\label{e27}
H_1=nt^{-1}+\frac{l_1}{3}t^{-3n}
\end{equation}
and
\begin{equation}
\label{e28}
H_3=nt^{-1}-\frac{2l_1}{3}t^{-3n}
\end{equation}
The mean Hubble parameter and deceleration parameter respectively are given as $H=\frac{n}{t}$ and $q=-1+\frac{1}{n}$.
Using the observational value for $q =-0.33\pm 0.17$ \cite{Kotambkar14}, we have restricted $n$ as $n\in(1.19,2)$ in case of polynomial law model. The scalar expansion and shear scalar are obtained as $\theta=\frac{3n}{t}$ and $\sigma^2=\frac{l_1^2}{3t^{6n}}$ respectively. For large values of cosmic time $t$, the scalar expansion and shear dies out.
\subsection{Constant scalar potential}
Let us consider here the constant scalar potential as \cite{Singh14}
\begin{equation}
\label{e29}
V(\phi)=V_0
\end{equation}
From equation \eqref{e20} and \eqref{e21} with the help of \eqref{e29}, we have obtained the cosmological constant and scalar field as
\begin{equation}
\label{e30}
\Lambda=-\frac{(4\lambda+1)V_0t^2+3n^2-n}{t^2}
\end{equation}
and
\begin{equation}
\label{e31}
\phi=-\frac{\sqrt{2}\left(\splitdfrac{\sqrt{-3n+t^{-6n+2}l_1^2}\sqrt{3}-}{3\sqrt{n}\arctan\left(\frac{\sqrt{-3n+t^{-6n+2}l_1^2}}{\sqrt{3n}}\right)}\right)}{\sqrt{2\lambda+1}\sqrt{\epsilon}(9n-3)}
\end{equation}
\begin{center}
\captionsetup{type=figure}
  \includegraphics[height=50mm]{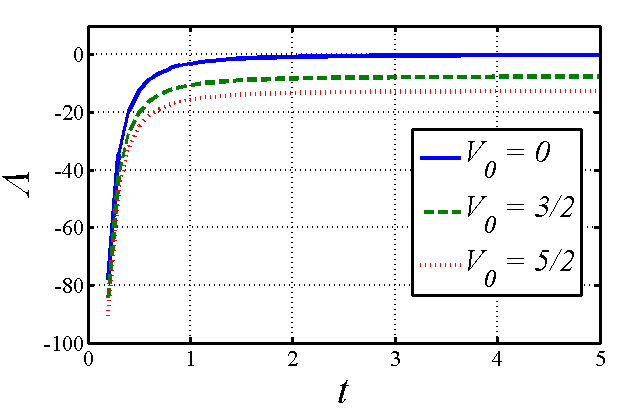}
  \caption{Variation of Cosmological constant $\Lambda$ against time $t$ for $n=1.2$, $\lambda=1$ and different $V_0(0, 3/2, 5/2)$ (power law).}
\label{fig1}
\end{center}

\begin{center}
\captionsetup{type=figure}
  \includegraphics[height=50mm]{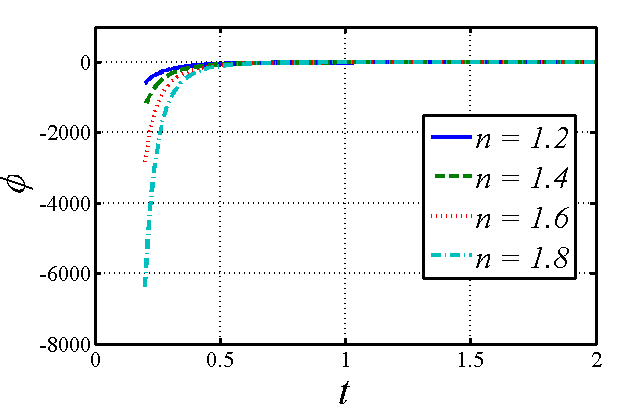}
  \caption{Variation of scalar field $\phi$ against time $t$ for $\lambda=1$, $\epsilon=1$, $l_1=50$ and different $n(1.2, 1.4, 1.6, 1.8)$ (power law).}
\label{fig2}
\end{center}

\begin{center}
\captionsetup{type=figure}
  \includegraphics[height=50mm]{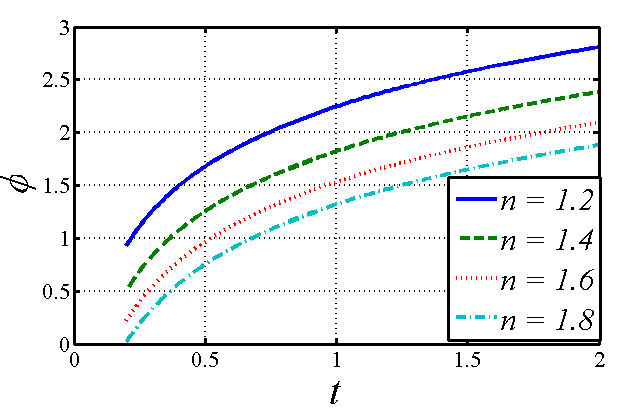}
  \caption{Variation of scalar field $\phi$ against time $t$ for $\lambda=1$, $\epsilon=-1$, $l_1=0.001$ and different $n(1.2, 1.4, 1.6, 1.8)$(power law).}
\label{fig3}
\end{center}

Fig.\ref{fig1} represents the variation of cosmological constant  $\Lambda$ against time for different parameters as in figure. Here, we have observed that $\epsilon$ does not affect the nature of the cosmological constant  $\Lambda$, which means it is independent of the normal (quintessence) and phantom scalar fields for the case of power law with constant potential. The cosmological constant is approaching towards zero with the evolution of time and also, we have observed that $\Lambda$ decreases with the increase of $V_0$.\\
Fig.\ref{fig2} and Fig.\ref{fig3} shows the scalar field $\phi$ against time for $\epsilon=\pm 1$ along with different parameters as in figures. Here, we have noticed that both scalar fields (normal (quintessence) and phantom scalar fields ) decreases with increase in $n$ and also, it is observed that normal (quintessence) scalar field is approaching towards zero with the evolution of time where as  phantom scalar field increases with the  evolution of time. It is noted that normal (quintessence) scalar field is negative and phantom scalar field is positive with the evolution of time.
\subsection{Exponential  scalar potential}

Let us consider here the exponential scalar potential as \cite{Burd88,Singh14}
\begin{equation}
\label{e32}
V(\phi)=V_0e^{-\beta \phi}
\end{equation}
where $V_0$ and $\beta$ are non-negative constants.\\
From equation \eqref{e20} and \eqref{e21} with the help of \eqref{e32}, we have obtained the  scalar potential and cosmological constant as
\begin{equation}
\label{e33}
V(t)=V_0exp\left(\dfrac{\beta\sqrt{2}\left(\splitdfrac{\sqrt{-3n+t^{-6n+2}l_1^2}\sqrt{3}}{-3\sqrt{n}\arctan\left(\frac{\sqrt{-3n+t^{-6n+2}l_1^2}}{\sqrt{3n}}\right)}\right)}{\sqrt{2\lambda+1}\sqrt{\epsilon}(9n-3)}\right)
\end{equation}
and
\begin{eqnarray}
\label{e34}
\Lambda(t)=\frac{1}{t^2}\bigg[-3n^2+n-t^2(4\lambda+1)V_0
exp\left(\dfrac{\beta\sqrt{2}\left(\splitdfrac{\sqrt{-3n+t^{-6n+2}l_1^2}\sqrt{3}}{-3\sqrt{n}\arctan\left(\frac{\sqrt{-3n+t^{-6n+2}l_1^2}}{\sqrt{3n}}\right)}\right)}{\sqrt{2\lambda+1}\sqrt{\epsilon}(9n-3)}\right)\bigg]\nonumber\\
\end{eqnarray}

\begin{center}
\captionsetup{type=figure}
  \includegraphics[height=50mm]{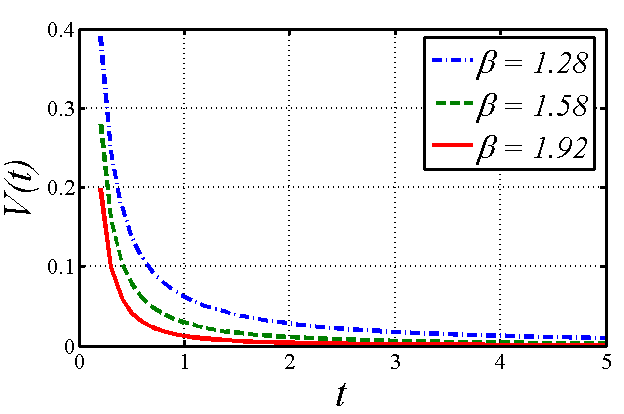}
\caption{Variation of scalar potential $V(t)$ against time $t$ for $\lambda=1$, $n=1.2$, $\epsilon=-1$,$l_1=0.001$, $V_0=3/2$ and different $\beta(1.28, 1.58, 1.92)$. We get the same result for $V_0=5/2$ along with other values as in the figure  (power law with exponential potential).}
\label{fig4}
\end{center}

\begin{center}
\captionsetup{type=figure}
  \includegraphics[height=50mm]{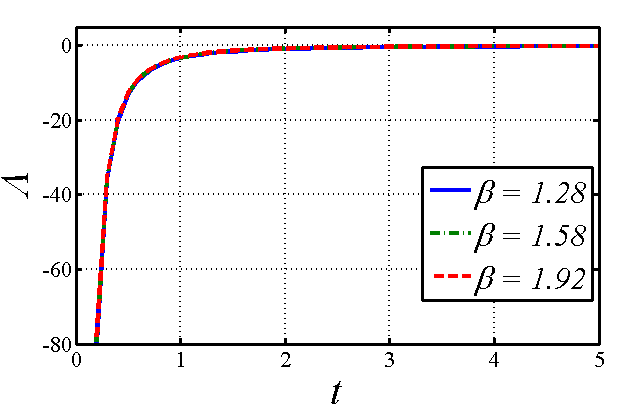}
\caption{Variation of cosmological constant $\Lambda$ against time $t$ for $\lambda=1$, $n=1.2$, $\epsilon=-1$, $l_1=0.001$, $V_0=3/2$ and different $\beta(1.28, 1.58, 1.92)$. We get the same result for $V_0=5/2$ along with other values as in the figure  (power law with exponential potential).}
\label{fig5}
\end{center}
Here, Fig.\ref{fig4} depicts the variation of scalar potential $V(t)$ against time for $\epsilon=-1$ and some consistent values $\beta = 1.28 \textrm{(SNe Ia)}, \beta = 1.58 \textrm{(H(z)+ SNe Ia)}\; \textrm{and}\;  \beta= 1.92 \textrm{(H(z)}$\cite{Singh14} along with other parameters as in figure. It is noticed from the figure that the higher the value of $\beta$, lower is the value of the scalar field potential and also the scalar field potential approaching towards zero with the evolution of time.  Fig.\ref{fig5} shows the variation of cosmological constant against time. It is observed that cosmological constant is  approaching towards zero with the evolution of time for $\epsilon=-1$. For $\epsilon=-1$, the scalar field potential and cosmological constant are complex valued hence, we neglect the case.

\section{Exponential Volume Expansion law Model }

We have considered the exponential volume expansion as
\begin{equation}
\label{e35}
V=e^{3nt}
\end{equation}
Using \eqref{e35} in \eqref{e22} and \eqref{e23} we get the scale factors as given by
\begin{equation}
\label{e36}
A=l_2^{1/3}exp\biggl[nt-\frac{l_1e^{-3nt}}{9n}\biggr]
\end{equation}
and
\begin{equation}
\label{e37}
B=l_2^{-2/3}exp\biggl[nt+\frac{2l_1e^{-3nt}}{9n}\biggr]
\end{equation}
The directional Hubble parameters are
\begin{equation}
\label{e38}
H_1=n+\frac{l_1}{3}e^{-3nt}
\end{equation}
and
\begin{equation}
\label{e39}
H_3=n-\frac{2l_1}{3}e^{-3nt}
\end{equation}
The Hubble parameter and deceleration parameter respectively are given as $H=n$ and $q=-1$. In this model, the expansion scalar  and shear scalar are obtained as $\theta=3n$ and $\sigma^2=\frac{l_1^2 e^{-6nt}}{3}$ respectively. At the initial epoch the  shear scalar becomes constant and when $t \rightarrow \infty$ the shear scalar approaches to zero.

\subsection{Constant scalar potential}

From equation  \eqref{e20} and \eqref{e21} with the help of \eqref{e29}, we have obtained the cosmological constant $\Lambda$ and scalar field $\phi$ as
\begin{equation}
\label{e40}
\Lambda=-(4V_0\lambda+V_0+3n^2)
\end{equation}
and
\begin{equation}
\label{e41}
\phi=-\frac{\sqrt{6}l_1e^{-3nt}}{9n\sqrt{(2\lambda+1)\epsilon}}
\end{equation}
\begin{center}
\captionsetup{type=figure}
  \includegraphics[height=50mm]{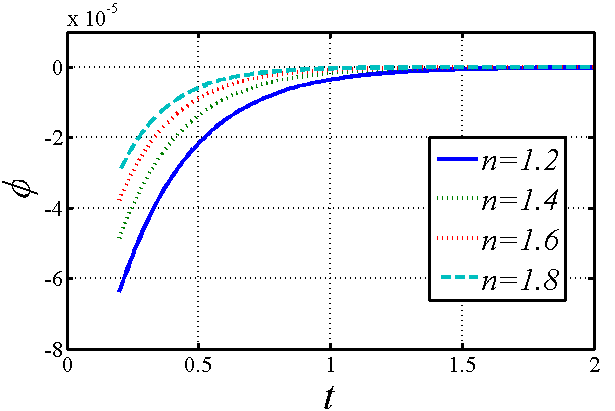}
\caption{Variation of scalar field $\phi$ against time $t$ for $\lambda=1$, $\epsilon=1$, $l_1=0.001$ and different $n(1.2, 1.4, 1.6, 1.8)$(exponential law).}
\label{fig6}
\end{center}
Fig.\ref{fig6} suggests that with the increase of $n$ the value of scalar field  increases, and it approaches towards zero with the evolution of time for $\epsilon=1$. We neglect the case $\epsilon=-1$ as it provides complex valued scalar field. In this case, the cosmological constant is independent of time and we have obtained that cosmological constant is a constant quantity (see eqn.\eqref{e40}).

\subsection{Exponential scalar potential}

From equation  \eqref{e20} and \eqref{e21} with the help of \eqref{e32}, we have obtained the  scalar field potential $V(t)$ and cosmological constant $\Lambda$ as
\begin{equation}
\label{e42}
V(t)=V_0exp\left(\frac{\beta\sqrt{6}l_1e^{-3nt}}{9n\sqrt{(2\lambda+1)\epsilon}}\right)
\end{equation}
and
\begin{equation}
\label{e43}
\Lambda=-(4\lambda+1)V_0exp\left(\frac{\beta\sqrt{6}l_1e^{-3nt}}{9n\sqrt{(2\lambda+1)\epsilon}}\right)-3n^2
\end{equation}
\begin{center}
\captionsetup{type=figure}
  \includegraphics[height=50mm]{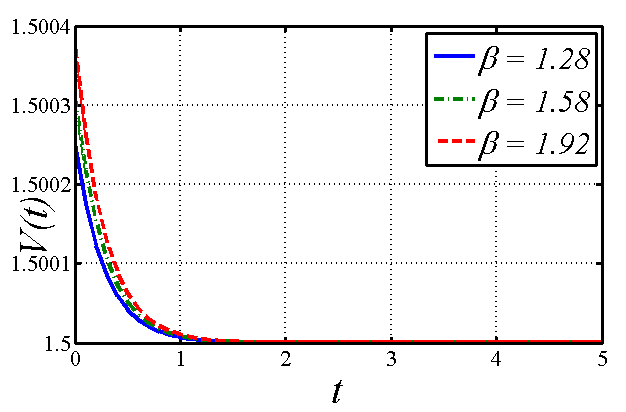}
\caption{Variation of scalar potential $V(t)$  against time $t$ for $\lambda=1$, $n=1.2$, $\epsilon=1$,$l_1=0.001$, $V_0=3/2$ and different $\beta(1.28, 1.58, 1.92)$.(exponential law with exponential potential)}
\label{fig7}
\end{center}

\begin{center}
\captionsetup{type=figure}
  \includegraphics[height=50mm]{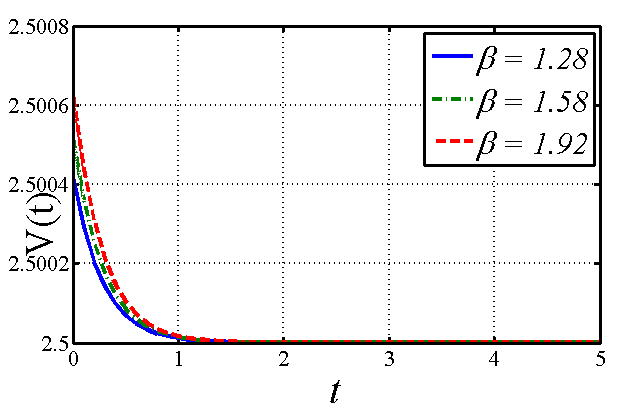}
\caption{Variation of scalar potential $V(t)$  against time $t$ for $\lambda=1$, $n=1.2$, $\epsilon=1$,$l_1=0.001$, $V_0=5/2$ and different $\beta(1.28, 1.58, 1.92)$  (exponential law with exponential potential).}
\label{fig8}
\end{center}

\begin{center}
\captionsetup{type=figure}
  \includegraphics[height=50mm]{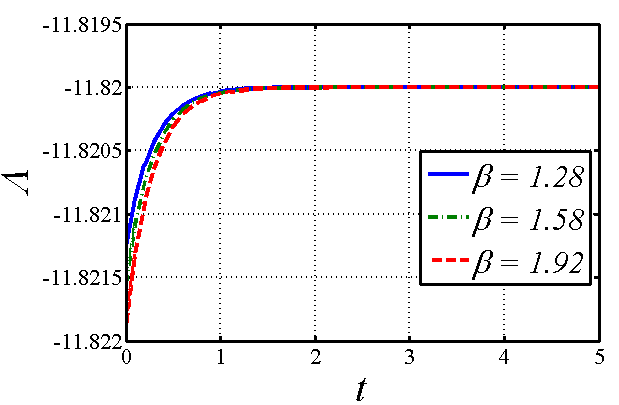}
\caption{Variation of cosmological constant $\Lambda$ against time $t$ for $\lambda=1$, $n=1.2$, $\epsilon=1$, $l_1=0.001$, $V_0=3/2$ and different $\beta(1.28, 1.58, 1.92)$  (exponential law with exponential potential).}
\label{fig9}
\end{center}

\begin{center}
\captionsetup{type=figure}
  \includegraphics[height=50mm]{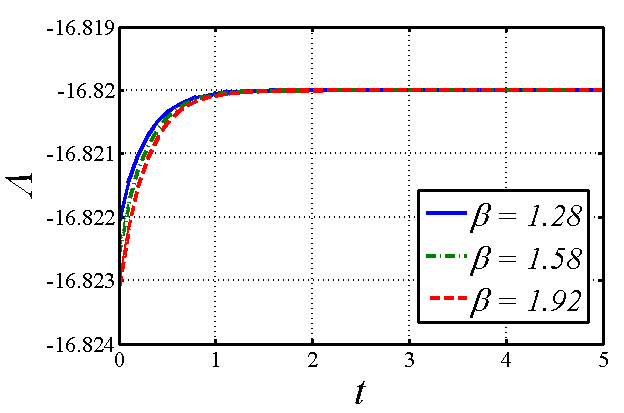}
\caption{Variation of cosmological constant $\Lambda$ against time $t$ for $\lambda=1$, $n=1.2$, $\epsilon=1$, $l_1=0.001$, $V_0=5/2$ and different $\beta(1.28, 1.58, 1.92)$  (exponential law with exponential potential).}
\label{fig10}
\end{center}
Fig.\ref{fig7} and Fig.\ref{fig8} have depicted the variation of scalar potential $V(t)$ against time for $\epsilon=1$ and some consistent values $ \beta = 1.28 \textrm{(SNe Ia)}, \beta = 1.58 \textrm{(H(z)+ SNe Ia)}\; \textrm{and}\;  \beta= 1.92 \textrm{(H(z)}$\cite{Singh14} along with other parameters as in figure. It is noticed that higher the value of $\beta$, higher becomes value of the scalar field potential (see Fig.\ref{fig7} and Fig.\ref{fig8}) and cosmological constant (see Fig.\ref{fig9} and Fig.\ref{fig10}). Also, here both scalar field potential and cosmological constant are approaching towards zero with the evolution of time.  The case $\epsilon=-1$, is neglected as it provides complex valued scalar field potential and cosmological constant.
\section{Conclusion}
In this article, we have studied the Bianchi type-I unverse in $f(R,T)$ gravity in presence of scalar field and cosmological constant. Here, we have discussed two models namely power law model and exponential law model. The observations obtained from these two models are presented below:
\begin{itemize}
  \item Power law volume expansion model.
  \begin{itemize}
  \item In case of power law volume expansion with constant potential, the cosmological constant $\Lambda$ is negative, normal (quintessence) and phantom scalar fields are negative and positive respectively.
  \item In case of power law volume expansion with exponential scalar potential, we have obtained same qualitative results as that of power law volume expansion with constant potential. In this case, the cosmological constant and scalar potential are dependent on scalar fields. For phantom scalar fields, cosmological constant and scalar potential are real (see Fig.\ref{fig4},Fig.\ref{fig5})
 \end{itemize}
  \item Exponential volume expansion law model.
  \begin{itemize}
  \item In case of exponential volume expansion law with constant potential, the cosmological constant $\Lambda$ is constant and  normal (quintessence)  scalar field is  negative.
  \item In case of exponential volume expansion law with exponential scalar potential, the cosmological constant $\Lambda$ is negative. Real cosmological constant and scalar field potential exist for only normal (quintessence) scalar field.
 \end{itemize}
\end{itemize}
Though, a lot of investigation has been done on $f(R,T)$ theory, still more research on this theory is required. Only one class of $f(R,T)$ gravity has been studied in this paper. This work can be extended to the other two  classes of $f(R,T)$ gravity and also the other modified theories of gravity.

\end{document}